\documentclass[aps,prd,twocolumn,eqsecnum,showpacs,amsmath]{revtex4}
\usepackage{graphicx}
\usepackage{amsfonts}
\usepackage{amssymb}

\newcommand{\dalm}{\kern1pt\vbox{\hrule height 0.9pt\hbox{\vrule width
0.9pt\hskip 2.5pt\vbox{\vskip 5.5pt}\hskip 3pt\vrule width 0.3pt}\hrule height
0.3pt}\kern1pt}

\begin{document}


\title{
Novel Cauchy-horizon instability}

\author{
Hideki Maeda $^{1}$
\footnote{Electronic address: hideki@gravity.phys.waseda.ac.jp}
,
Takashi Torii $^{2}$
\footnote{Electronic address: torii@gravity.phys.waseda.ac.jp}
, and
Tomohiro Harada $^{3}$
\footnote{Electronic address: T.Harada@qmul.ac.uk}
}

\address{ 
$^{1}$Advanced Research Institute for Science and Engineering,
Waseda University, Okubo 3-4-1, Shinjuku, Tokyo 169-8555, Japan\\
$^{2}$ Graduate School of Science,
Waseda University, Okubo 3-4-1, Shinjuku, Tokyo 169-8555, Japan\\
$^{3}$Astronomy Unit, School of Mathematical Sciences,
Queen Mary, University of London,
Mile End Road, London E1 4NS, UK
}

\date{\today}

\begin{abstract}                
The evolution of weak discontinuity is 
investigated on horizons in the $n$-dimensional static
solutions in the Einstein-Maxwell-scalar-$\Lambda$ system,
including the Reissner-Nordstr\"om-(anti) de Sitter black hole.
The analysis is essentially local and nonlinear.
We find that 
the Cauchy horizon is unstable, whereas both
the black-hole event horizon and the cosmological 
event horizon are stable.
This new instability, the so-called kink instability, 
of the Cauchy horizon is completely different
from the well-known ``infinite-blueshift'' instability.
The kink instability makes
the analytic continuation beyond the Cauchy horizon unstable.
\end{abstract}

\pacs{04.20.Dw, 04.70.-s, 04.40.Nr, 04.50.+h} 

\maketitle

\section{Introduction}
A ``horizon'' generally means a surface across which no information
can pass.
In gravitational physics, there are two important horizons, 
the {\em event horizon} and the {\em Cauchy horizon}.
When the gravitational field is very strong,
some light rays that are emitted outwardly may
run inwardly later.
The event horizon is traced out by critical
light rays which escape to infinity. 
The region surrounded by the event
horizon is called a black hole.  
The stability of the black hole
against the perturbations outside the event horizon has been well
investigated~\cite{chandrasekhar}. 
On the other hand, the Cauchy horizon is the future
boundary of the Cauchy development 
of a partial Cauchy surface
so that the predictability of physics breaks 
down beyond the Cauchy horizon. 
See Ref.~\cite{he1973} for
a rigorous definition of the event horizon
and the Cauchy horizon.

It was subsequently proved
that spacetime singularities inevitably appear under 
general situations and physical energy conditions~\cite{he1973}. 
Gravitational collapse is one of the physical processes where
singularities must appear. In this context, a {\em cosmic censorship
hypothesis} (CCH) was proposed by Penrose, which asserts that 
singularities formed in generic gravitational collapse of physical
matters cannot be observed;
in other words, there are no {\em naked singularities} formed 
in physical gravitational collapse~\cite{penrose1969,penrose1979}.
The Reissner-Nordstr\"om (RN) solution and the 
Reissner-Nordstr\"om-de Sitter (RNdS) solution 
have a Cauchy horizon as an inner horizon. 
Inside it there is a central timelike singularity. 
However, Penrose demonstrated that perturbations 
originating outside the black hole would
be blue-shifted infinitely
at the Cauchy horizon,
which results in a ``blue-sheet'' singularity~\cite{penrose1968}. 
It was found 
that this Cauchy horizon is
unstable against perturbations and
transforms into a null weak curvature singularity~\cite{pi1990,ori1991,burko1997,burko2003,brady1999}. 
In the presence of a positive cosmological constant, i.e., the RNdS solution, 
the Cauchy horizon is also unstable~\cite{chambers1997,bmm1998}. 

The general proof of CCH is, however, far from complete, and many
counterexample candidates have been proposed in the framework of
general relativity~\cite{harada2004}. Understanding the stability of 
naked-singular solutions gives clearly 
significant insight into the issue of CCH, and
further analyses which involve the full non-linear
perturbations are required. 

In this paper we study the stability of horizons against 
perturbations 
for a large class of static solutions. 
We consider the Einstein-Maxwell-scalar-$\Lambda$ system in the
$n$-dimension, which contains the Reissner-Nordstr\"om-(anti) de Sitter
(RN(A)dS) solution as a special solution.
In recent years, higher-dimensional asymptotically anti-de Sitter
spacetimes with spherical, plane, or hyperbolic symmetry play an
important role as a bulk spacetime in 
the brane-world scenario~\cite{brane}, 
so they are included in the analysis.  
The perturbation analysis 
is nonlinear and full order.
This kind of perturbation is called a kink-mode perturbation. 
Similar stability analyses 
have been done in spherically symmetric self-similar solutions
in Newtonian gravity~\cite{op1988,hm2000}
and general relativity~\cite{harada2001,hm2004,mh2004} and in the context of the stationary accretion-disk flow to a compact object~\cite{accretion}.

The organization of this paper is the following.
In Section II, basic equations are presented.
In Section III, the stability for the kink mode is analyzed in full order, and a stability criterion for this mode is derived.
In Section IV, applications to known static solutions, such as the RN(A)dS solution, are presented and the implications of the kink instability of the Cauchy horizon are discussed.
We adopt units such that $c=1$.

\section{Model and background static solutions}
We begin with the following $n$-dimensional Einstein-Maxwell-scalar-$\Lambda$
system with the action:
\begin{eqnarray}
\label{action}
&& S=\int
d^nx\sqrt{-g}\biggl[\frac{1}{2\kappa_n^2}\bigl(R-2\Lambda\bigr)
-\frac12\nabla_\mu\phi\nabla^\mu\phi-V(\phi)
\nonumber
\\ 
&& \;\;\;\;\;\;\;\;\;\;\;\;\;\;\;\;\;\;\;\;\;\;\;\;\;\;\;\;\;\;\;\;
-\frac{1}{4\pi g_c^2}{ F}_{\mu\nu}{ F}^{\mu\nu}\biggr],
\end{eqnarray}
where $R$ and $\Lambda$ are the $n$-dimensional scalar curvature 
and the cosmological
constant, respectively.
$\kappa_n:=\sqrt{8\pi G_n}$, where $G_n$ is the gravitational constant in the $n$-dimension and $g_c$ is the gauge coupling constant of the Maxwell
field $F^{\mu\nu}$.

The basic equations are given by 
\begin{eqnarray}
R_{\mu\nu}-\frac{2}{n-2} \Lambda g_{\mu\nu}&=&\kappa_n^2 \Bigl[S^{(
A)}_{\mu\nu}+S^{(\phi)}_{\mu\nu}\Bigr],
\label{einstein}\\
\nabla_{\nu}F^{\mu\nu}&=&0, 
\label{maxwell}\\
\dalm\phi&=&\frac{d V}{d \phi},\label{klein-gordon}
\end{eqnarray}  
where
\begin{eqnarray}
&&
S^{(A)}_{\mu\nu}:=\frac{1}{4\pi
g_c^2}\biggl[F_{\mu\sigma}F_{\nu}^{~\sigma}-\frac{1}{2(n-2)}g_{\mu\nu}F_{\rho\sigma}F^{\rho\sigma}\biggl],
\nonumber
\\
\\
&&
S^{(\phi)}_{\mu\nu}:=\partial_\mu\phi\partial_\nu\phi +\frac{2}{n-2}g_{\mu\nu}V(\phi).
\end{eqnarray}  
We consider $n$-dimensional spacetimes with the metric
\begin{equation}
\label{metric}
ds^2=-fe^{2A}dv^2+2e^{A}dvdr+r^2d\Omega_{n-2}^2,
\end{equation}  
where $f=f(v,r)$, $A=A(v,r)$, and $d\Omega_{n-2}^2=\gamma_{ij}dx^i dx^j~(i,j=2\cdots n)$ is the metric of the $(n-2)$-dimensional
unit Einstein space, which includes the $(n-2)$-dimensional unit sphere,
plane and hyperboloid.
$v$ is the advanced time coordinate so that 
a curve $v=\mbox{const}$ denotes a radial ingoing null geodesic.
Throughout this paper, we call the region with the smaller (larger) value of $r$ ``inside (outside) region.''

The Maxwell equations
(\ref{maxwell}) can be easily integrated to give 
\begin{eqnarray}
\label{elemag}
F_{\mu\nu}dx^\mu \wedge dx^\nu=\frac{Q}{r^{n-2}}e^{A}dv\wedge dr,
\end{eqnarray}  
where we only consider the electric field produced by a constant charge $Q$. Then the Einstein equations and the equation
of motion for the scalar field reduce to the following partial differential equations: 
\begin{eqnarray}
-\frac{n-2}{2r}{\dot f}e^A
=\kappa_n^2\bigl({\dot \phi}^2+fe^A{\dot\phi}\phi'\bigr),
\label{basic2}
\end{eqnarray}  
\begin{eqnarray}
\frac{n-2}{r}A'=\kappa^2_n\phi^{'2},
\label{basic3}
\end{eqnarray}  
\begin{eqnarray}
&&
(n-3)(k-f)-r(fA'+f')-\frac{2\Lambda}{n-2} r^2
\nonumber 
\\ 
&& \;\;\;\;\;
=\frac{\kappa^2_n}{n-2}\biggl[2r^2V+\frac{1}{4\pi
g_c^2}\frac{Q^2}{r^{2(n-3)}}\biggl],
\label{basic4}
\end{eqnarray}  
\begin{eqnarray}
&&f\phi''+e^{-A}\left(2{\dot\phi}'+\frac{n-2}{r}{\dot\phi}\right)+\biggl(\frac{f'}{f}+A'+\frac{n-2}{r}\biggl)f\phi' 
\nonumber 
\\
&& \;\;\;\;\;\;\;
=\frac{d V}{d \phi},
\label{basic5}
\end{eqnarray}  
where a dot and a prime denote the partial derivatives with respect to $v$ and $r$,
respectively. 
Three of the above four equations are independent. 
$k=0,\pm 1$ denotes the curvature of the $(n-2)$-dimensional submanifold.

In the static case, we assume that $f=f(r),A=A(r)$, and
$\phi=\phi(r)$, and drop the terms with dots. Then 
Eqs.~(\ref{basic2})--(\ref{basic5})
reduce to a  set of ordinary differential equations. 
These equations are singular at $r=r_{\rm h}$, where $f(r_{\rm h})=0$. 
Along a future-directed outgoing null geodesic, we have $dr/dv=f/2$.
Therefore, $r=r_{\rm h}$ is a future-directed null geodesic
and a horizon. The region with $f<0$ is trapped.
No information comes out from the trapped region across the horizon.
The horizon is termed as $transluminal$, $anti$-$transluminal$
, and $degenerate$ for $f'>0$, $<0$, and $=0$, respectively.

\begin{figure}[tbp]
\includegraphics[width=.80\linewidth]{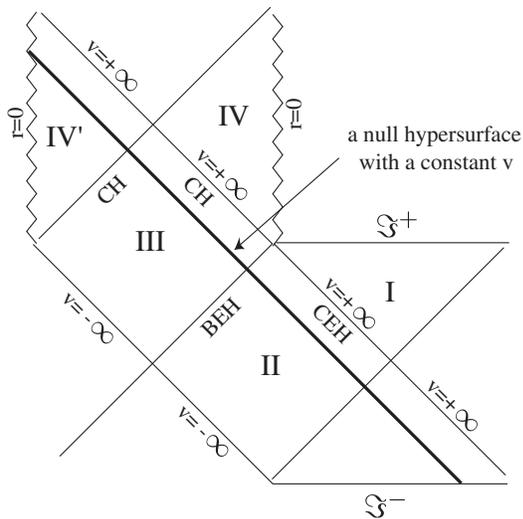}
\caption{
A portion of the conformal diagram for the maximally-extended RNdS solution with three horizons.
Zigzag lines represent the timelike central singularities. $\Im^{+(-)}$ corresponds to the future (past) null infinity. The black-hole event horizon (BEH) is the null hypersurface between Regions II and III. 
The inner horizon as a Cauchy horizon (CH) is that between Regions III and IV or IV$'$. 
The cosmological event horizon (CEH) is that between Regions I and II. 
Regions I, II, and III correspond to the exterior of the cosmological event horizon, the Region between the black-hole event horizon and the cosmological event horizon, and the Region between the inner horizon and the black-hole event horizon, respectively. 
Regions IV and IV$'$ correspond to the interior of the inner horizon. 
}
\label{RNdS}
\end{figure}

We consider static solutions with at least
one regular horizon. By the condition of the regularity
of horizon, functions $f$, $A$, $\phi$ and $\phi'$ and their derivatives 
with respect to
$r$ are finite at the horizon. 
An example of the spherically symmetric static solution with analytic
horizons is the $n$-dimensional RN(A)dS solution, which is given by 
\begin{equation}
\label{rn}
f=1-\frac{\tilde{M}}{r^{n-3}}+\frac{(n-3)}{2}\frac{\tilde{Q}^2}{r^{2n-6}}
-\frac{2\Lambda_{\rm eff}}{(n-1)(n-2)}r^2,
\end{equation}  
\begin{equation}
\Lambda_{\rm eff} := \Lambda+\kappa_n^2V(\phi_0),  \;\;\;\;
A\equiv 0,  \;\;\;\;
\phi \equiv \phi_0,
\end{equation}  
where $\tilde{M}$ and  $\tilde{Q}$ are constants related to the mass $M$ and
the charge $Q$ of the black hole as
\begin{equation}
\tilde{M}:=\frac{16\pi G_n M}{(n-2)\!\Sigma_{n-2}},
\;\;\;\;\;
\tilde{Q}^2:=\frac{\kappa_n^2 Q^2}{2(n-2)\pi g_c^2},
\end{equation}  
respectively,
where $\Sigma_{n-2}$ is the volume of the 
$(n-2)$-dimensional unit sphere.
$\phi_0$ is the
extremal point of the potential $V$. In this spacetime, the number of
horizons varies depending on
$\tilde{M}$, $\tilde{Q}$, and $\Lambda_{\rm eff}$ and is always 
less than or equal to three.
The RNdS spacetime with three horizons has, as shown in Fig.~\ref{RNdS},
a Cauchy horizon, a black-hole event horizon and a cosmological event horizon.
There are other known black-hole solutions with analytic horizons, in which the effective cosmological
constant $\Lambda_{\rm eff}$ is negative and the first term of the 
right-hand side in Eq.~(\ref{rn}) is replaced by $0$ or $-1$.
They are called topological black holes, since the topology
of the constant $r$ surface is not $S^{n-2}$.
The spacetime structure of these solutions 
has been investigated~\cite{Brill,tm2004}. Spherically symmetric static black-hole solutions
with the nontrivial configuration of a scalar field, i.e., a scalar hair,
with a double-well potential, has been also found numerically
both in four-dimensional asymptotically de Sitter and
anti-de Sitter spacetimes, where the former is unstable against
spherical perturbations while the latter is stable~\cite{tmn}.

\section{Kink instability of horizons}
We consider full-order radial perturbations $\delta
f(v,r),\delta A(v,r)$, and $\delta \phi(v,r)$ in general background static solutions
such as
\begin{eqnarray}
f(v,r)&=&f_{(\rm b)}(r)e^{\delta f(v,r)},\\
A(v,r)&=&A_{(\rm b)}(r)+\delta A(v,r),\\
\phi(v,r)&=&\phi_{(\rm b)}(r)+\delta \phi(v,r),
\end{eqnarray}  
where $f_{(\rm b)},A_{(\rm b)}$, and $\phi_{(\rm b)}$ denote a background
static solution that satisfies Eqs.~(\ref{basic2})--(\ref{basic5}). We consider perturbations that satisfy the following conditions: (i) the initial perturbations only
exist inside the horizon, i.e., $r<r_h$;
(ii) $f$, $A$, $\phi$, and $\phi'$ are continuous
at the horizon; (iii) $\phi''$ and $\dot{\phi}''$ are
discontinuous at the horizon
although they have finite one-sided limit values as $r\to r_{\rm h}-0$ and $r\to
r_{\rm h}+0$; (iv) no curvature singularities exist at the horizon at the initial moment.

Now we consider the behavior of perturbations at the horizon.
By the condition (i), the region outside the horizon remains
unperturbed, i.e., it is described by the background static solution, because no information can propagate beyond the horizon.
By condition (ii), $\delta A$, $\delta f$, $\delta \phi$, and 
$\delta \phi'$ vanish at the horizon so that 
\begin{eqnarray}
\delta A'=0 \label{deltaA'}
\end{eqnarray}
is satisfied at the horizon from Eq.~(\ref{basic3}). From Eqs.~(\ref{basic4}) and (\ref{deltaA'}), 
\begin{eqnarray}
\lim_{r\to r_{\rm h}}f_{(\rm b)}\delta f'=0 \label{fdeltaf'}
\end{eqnarray}
is obtained. Differentiating Eq.~(\ref{basic3}) with respect to $r$ and 
evaluating both sides at
the horizon, we obtain 
\begin{eqnarray}
\frac{n-2}{2r_{\rm h}}\delta A''=\kappa^2_n \phi_{(\rm b)}'\delta \phi'',\label{deltaA''}
\end{eqnarray}  
which means that $\delta A''$ is finite because of condition
(iii). Differentiating Eq.~(\ref{basic4}) with respect to $r$, 
evaluating both sides at
the horizon, and 
using condition (iii) and Eqs.~(\ref{deltaA'})--(\ref{deltaA''}), we obtain 
\begin{eqnarray}
\lim_{r\to r_{\rm h}}[2f_{(\rm b)}'\delta f'+f_{(\rm b)}(\delta f''+\delta f^{'2})]=0.\label{key1}
\end{eqnarray}  
Differentiating Eq.~(\ref{basic5}) with respect to $r$, and using condition (iii) and Eqs.~(\ref{deltaA'})--(\ref{key1}), we find 
\begin{equation}
\label{key}
2e^{-A_{(\rm b)}}\delta\dot{\phi}''+2f_{(\rm b)}'\delta\phi''
+\lim_{r\to r_{\rm h}}f_{(\rm b)}\delta\phi'''=0
\end{equation}  
at the horizon. 
We can prove the third term on the left-hand side vanishes in
the same way as in Appendix~B in Ref.~\cite{hm2004}. As a result, Eq.~(\ref{key}) becomes
\begin{equation}
\delta{\dot \phi}'' = -f_{(\rm b)}'e^{A_{(\rm b)}}\delta\phi''.
\end{equation}  
It should be noted that the perturbations are of full-order although this equation is linear. This is integrated to give 
\begin{equation}
\label{solution}
\delta {\phi}'' \propto e^{-\alpha v},
\end{equation}  
where
\begin{eqnarray}
\alpha:=f_{(\rm b)}'e^{A_{(\rm b)}}.
\end{eqnarray}  

Here we define instability by the exponential growth of discontinuity.
Then we find the following criterion:  
anti-transluminal horizons,
i.e., horizons with $f'>0$, are stable,
while transluminal horizons, i.e., horizons with $f'<0$, are
unstable. Degenerate horizons, i.e., horizons with
$f'=0$, are marginally stable.
Irrespective of the
form of the potential $V(\phi)$, this stability criterion
applies to any analytical and numerical solutions
with regular horizons. 
In the above, we have discussed the stability of horizons 
that are future-directed outgoing null geodesics.
If horizons are given by future-directed ingoing null geodesics,
such as a Cauchy horizon of RN(A)dS solution,
we find serious coordinate degeneracy in the $(v,r)$ coordinates.
Hence, we need to reformulate the perturbation analysis 
using the $(u,r)$ coordinates:
\begin{equation}
ds^2=-fe^{2A}du^2-2e^{A}dudr+r^2d\Omega_{n-2}^2,
\end{equation}
where $f=f(u,r)$ and $A=A(u,r)$. 
$u$ is the retarded time coordinate. Since the above form can be obtained 
from Eq.~(\ref{metric}) through the coordinate transformation $v=-u$,
the stability in terms of $u$ should be reversed 
from that in terms of $v$.
Therefore, the stability of horizons which are given by 
future-directed ingoing null geodesics are the following:
anti-transluminal horizons are unstable,
while transluminal horizons are
stable. Degenerate horizons are marginally stable.
In Appendix~\ref{gauge}, we demonstrate in the four-dimensional spherically symmetric spacetime that these perturbations are gauge-invariant up to linear order.

\section{Discussion}
We have obtained the stability of horizons in $n$-dimensional static solutions in the
Einstein-Maxwell-scalar-$\Lambda$ system without assuming the explicit form of
the potential of the scalar field. The RN(A)dS
solution is included in the analysis as a special case. 
The present
work has shown for the first time the existence of the kink instability of
horizons in static solutions in general relativity. The intriguing
feature is that the kink instability grows exponentially in terms of $v$.
In contrast the kink mode perturbation blows up to infinity in a finite
time in the self-similar perfect-fluid system with
$p=k\mu$ ($0<k<1$) in general relativity~\cite{harada2001}.

The stability of solutions against kink-mode perturbation is determined
only by the local
property, i.e., the sign of $f'$ at the horizon of the background 
static solutions.
This criterion applies to a large class of horizons. 
For the RNdS spacetime, a black-hole event
horizon
is stable against the kink-mode perturbation. A
cosmological event horizon
is also stable. On the other hand, a Cauchy horizon 
is unstable. 
As a result, a wide class
of solutions 
suffers from the kink instability. 
There is a significant difference here from the case of
self-similar solutions where
the black-hole event horizon is stable
but the Cauchy horizon and the cosmological
event horizon are not always unstable~\cite{harada2001,hm2004}.

We consider the implications of the kink instability of horizons. The
initial perturbations of the kink mode are inserted only in the future of
the horizons,
and therefore, this instability does $not$ prevent the horizons or the
naked singularities from forming. The kink instability implies that, if the
analyticity of a horizon is violated even weakly, the perturbed spacetime
is much different from that with an analytic horizon. Even so, this kind of instability does not
cause the formation of a singularity on the horizon in finite time because the growth 
of the discontinuity is only exponential.

An important example of kink-unstable horizons is the Cauchy horizon of 
the RN(A)dS solution, which associates with a timelike naked
singularity. It is generated by the first future-directed null ray 
emanated from the naked singularity. If the naked singularity forms in a gravitational collapse, the information from the singularity, which is physically unpredictable, affects only the future of the Cauchy horizon, and the naked singularity may violate its analyticity. Then, the inside region of the Cauchy horizon is represented by a spacetime much different from that in the RN(A)dS solution. 

As a thought experiment, suppose that one drops an ideal small apparatus
into the RN(A)dS black hole. 
In four-dimensional spherically symmetric case, the apparatus never arrives at the regular Cauchy horizon because its back reaction to the spacetime actually disturbs the background before it reaches the Cauchy horizon and the Cauchy horizon is transformed into a null weak curvature singularity~\cite{pi1990,ori1991,burko1997,burko2003,brady1999,chambers1997,bmm1998}.
Here we assume that the ideal apparatus has so small mass that its back reaction to the spacetime can be negligible but can disturb the scalar field at any time.
Then, it falls across the event horizon and reaches the Cauchy horizon. 
At this moment, if it disturbs the scalar field, so that 
the second derivative of the scalar field is discontinuous, 
the disturbance grows up through kink instability 
and propagates along the Cauchy horizon at the speed of light. 
As a result, the gravitational field inside of the 
Cauchy horizon is very much modified.


\acknowledgments
HM would like to thank Shoji Kato and Ken-ichi Nakao for useful comments. This work was partially supported by a Grant for The 21st Century COE Program (Holistic Research and Education Center for Physics Self-Organization Systems) at Waseda University. TH was supported from the JSPS.

\appendix

\section{gauge-invariance of kink instability}
\label{gauge}
We demonstrate in the four-dimensional spherically symmetric spacetime that the perturbations are gauge-invariant up to linear order.
All notations here follow~\cite{gs}.
We write the spherically symmetric spacetime as a product manifold ${\cal M}=M^2\times S^2$ with metric
\begin{equation}
g_{\mu\nu}=\mbox{diag}(g_{AB},r^2\gamma_{ab}),
\end{equation}
where $g_{AB}$ is an arbitrary Lorentz metric on $M^2$, $r$ is a scalar on $M^2$ with $r=0$ defining the boundary of $M^2$, and $\gamma_{ab}$ is the unit curvature metric on $S^2$. We introduce the covariant derivatives on spacetime ${\cal M}$, the subspacetime $M^2$ and the unit sphere $S^2$ with
\begin{eqnarray}
g_{\mu\nu;\lambda}&=&0,\\
g_{AB|C}&=&0,\\
g_{ab:c}&=&0.
\end{eqnarray}
We denote a perturbed scalar field $\phi$ as
\begin{equation}
\phi=\phi_{(\rm b)}+\delta \phi.
\end{equation}
Hereafter we linearize the perturbation.

The perturbation of the scalar field is transformed by a gauge transformation of even parity as
\begin{eqnarray}
\delta {\bar \phi}=\delta \phi-\phi_{{(\rm b)};\mu}\xi^\mu,
\end{eqnarray}
where $\xi^\mu$ is a generator of the infinitesimal coordinate transformation.

Spherically symmetric metric perturbations are written as
\begin{eqnarray}
\delta g_{AB}&=&h_{AB}Y, \\
\delta g_{ab}&=&r^2 KY\gamma_{ab},
\end{eqnarray}
where $h_{AB}$ and $K$ are a tensor and a scalar on $M^2$ and $Y$ is a constant.
For spherical perturbations, all gauge transformations we have for metric perturbations are
\begin{eqnarray}
{\bar h}_{AB}&=&h_{AB}-(\xi_{A|B}+\xi_{B|A}), \\
{\bar K}&=&K-2v_A\xi^A,\label{ap1}
\end{eqnarray}
where $v_A:=r_{,A}/r$.
Here, for simplicity, we have chosen the areal coordinate $r$ as a radial coordinate, i.e., $x^1=r$, for the background spacetime.

Since
\begin{equation}
v_A=\frac{r_{|A}}{r}=\left(0,\frac{1}{r}\right),
\end{equation}
we have
\begin{equation}
\xi^1=-\frac{r}{2}({\bar K}-K),
\end{equation}
from Eq.~(\ref{ap1}).
Using the fact that $\phi_{(\rm b)}$ depends only on $r$, we can construct the following gauge invariant perturbation $\Phi$:
\begin{eqnarray}
\Phi=\delta \phi-\frac{r}{2}K\frac{d\phi_{(\rm b)}}{dr}.
\end{eqnarray}
When we choose the areal coordinate $r$ as a radial coordinate in the perturbed spacetime, which is adopted in the main text of the present article, $K=0$ is satisfied so that the gauge invariant quantity $\Phi$ is identical to $\delta \phi$ in this gauge choice.
For the gradient and higher derivatives of the scalar field, we can construct them just by covariantly differentiating $\Phi$ on the background metrics.


\end{document}